\documentclass{aa}
\title{[CII] emission properties of the massive
star-forming region RCW~36 in a filamentary molecular cloud}
\titlerunning{[CII] observations of RCW~36}

\usepackage{color}
\usepackage{natbib}

\bibpunct{(}{)}{;}{a}{}{,} 
\authorrunning{Suzuki~et~al.}
\author{T.~Suzuki\inst{\ref{inst1}}, S.~Oyabu\inst{\ref{inst1}}, 
S.~K.~Ghosh\inst{\ref{inst2}}, D.~K.~Ojha\inst{\ref{inst2}}, H.~Kaneda\inst{\ref{inst1}},
H.~Maeda\inst{\ref{inst1}}, T.~Nakagawa\inst{\ref{inst3}}, J.~P.~Ninan\inst{\ref{inst4}},
S.~Vig\inst{\ref{inst5}}, M.~Hanaoka\inst{\ref{inst1}}, F.~Saito\inst{\ref{inst1}},
S.~Fujiwara\inst{\ref{inst1}}, and T.~Kanayama\inst{\ref{inst1}}}
\institute{Graduate School of Science, Nagoya University, Furo-cho, Chikusa-ku, Nagoya, Aichi, 464-8602, Japan\label{inst1}
\and 
Tata Institute of Fundamental Research, 
Homi Bhabha Road, Colaba, Mumbai 400005, India\label{inst2}
\and 
Institute of Space and Astronautical Science, Japan Aerospace
Exploration Agency 3-1-1 Yoshinodai, Chuo-ku, Sagamihara, Kanagawa, 252-5210,
Japan\label{inst3}
\and The Pennsylvania State University, University Park, State College, PA, USA\label{inst4}
\and Indian Institute of Space Science and Technology, Valiamala,
Thiruvananthapuram 695 547, India\label{inst5}}
\date{Received / Accepted}
\abstract{}{To investigate properties of [\ion{C}{ii}]~158~$\mu$m
emission of RCW~36 in a dense filamentary cloud.}{[\ion{C}{ii}]
observations of RCW~36 covering an area of $\sim30 \arcmin \times 30\arcmin$
were carried out with a Fabry-P\'{e}rot spectrometer aboard a 100-cm
balloon-borne far-infrared (IR) telescope with an angular resolution of
90\arcsec. By using AKARI and \textit{Herschel} images, the spatial distribution
of the [\ion{C}{ii}] intensity was compared with those of emission from
the large grains and polycyclic aromatic hydrocarbon (PAH).}{The
[\ion{C}{ii}] emission is spatially in good agreement with shell-like
structures of a bipolar lobe observed in IR images, which extend along the
direction perpendicular to the direction of a cold dense
filament. We found that the [\ion{C}{ii}]--160~$\mu$m
relation for RCW~36 shows higher brightness ratio of
[\ion{C}{ii}]/160~$\mu$m than that for RCW 38, while the
[\ion{C}{ii}]--9~$\mu$m relation for RCW~36 is in good agreement with
that for RCW~38.}{By spectral
decomposition analysis on a pixel-by-pixel basis 
using IR images, the [\ion{C}{ii}] emission spatially well correlates
with PAH and cold dust emissions. This means that the observed [\ion{C}{ii}]
emission dominantly comes from PDRs. Moreover, the
$L_\mathrm{[\ion{C}{ii}]}/L_\mathrm{FIR}$ ratio shows large variation 
($10^{-2}$--$10^{-3}$) compared with the
$L_\mathrm{[\ion{C}{ii}]}$/$L_\mathrm{PAH}$ ratio. In view of the
observed tight correlation between
$L_\mathrm{[\ion{C}{ii}]}/L_\mathrm{FIR}$ and the optical depth at
$\lambda=160$~$\mu$m, the large variation in
$L_\mathrm{[\ion{C}{ii}]}/L_\mathrm{FIR}$ can be simply explained by the
geometrical effect, viz., $L_\mathrm{FIR}$ has contributions from
the entire dust-cloud column along the line of sight, while
$L_\mathrm{[\ion{C}{ii}]}$ has contributions from far-UV
illuminated cloud surfaces. Based on the picture
of the geometry effect, the enhanced brightness ratio of
[\ion{C}{ii}]/160~$\mu$m is attributed to the difference in gas
structures where massive stars are formed: filamentary (RCW~36) and
clumpy (RCW~38) molecular clouds and thus suggests
that RCW~36 is dominated by far-UV illuminated cloud surfaces compared with RCW~38.} 

\begin{document}
\maketitle

\section{Introduction}
Understanding of galaxy evolution remains a key subject in
modern astrophysics. Massive stars ($\ga 8~M_\sun$) have a great
influence on the evolution of the interstellar medium (ISM)
energetically, chemically, and dynamically, and thus plays a vital role 
in galaxy evolution. More recently, \textit{Herschel} observations
reveal the presence of ubiquitous filamentary structures connecting Galactic
star-forming regions with each other, suggesting that filaments have the
crucial role on star formation~\citep[e.g.][]{Molinari2010,
Andre2010}. Special attention on filament structures has been 
promoting investigations of the radiative feedback of massive stars on
their filamentary ISM, which regulates subsequent star formation as well
as the role of filaments in the formation of dense star-forming
clumps~\citep{Zinnecker2007,Myers2009, Andre2010,
Baug2015, Dewangan2017a, Dewangan2017b}.

A photo-dissociation region (PDR) is formed around an
\ion{H}{ii} region by massive stars, and is the
region where far-ultraviolet (UV)~($6<h \nu<13.6$~eV) photons play a
significant role in the heating and chemistry of the ISM; all of the
atomic and at least 90\% of the molecular gas in the Milky Way are in
PDRs~\citep{Hollenbach1999}. 
This means that PDRs are key to understanding the physical properties of
the ISM strongly influenced by the radiative feedback of massive stars,
and the triggering of next-generation stars. In PDRs, the gas heating is
mainly caused by photoelectrons (photoelectric heating) from dust grains
and polycyclic aromatic hydrocarbons~(PAHs), which are ejected by
absorbing far-UV photons from stars, while the gas cooling is performed
by far-infrared~(IR) fine-structure lines, among which
[\ion{C}{ii}]~158~$\mu$m is the most important gas coolant in
low-density PDRs~\citep[e.g.][]{Hollenbach1999}.

Studies of gas cooling and photo-electric heating of gas
in PDRs require [\ion{C}{ii}] and IR photometric data sets and sufficient
spatial resolution to trace the variety of dust emission components such
as PAHs, warm ($\sim$60--100~K), and cold ($\sim$20--30~K ) dust emissions and to
diagnose region-by-region physical conditions in the ISM. The reason is
that gas heating and cooling processes depend on dust grain types and
ISM conditions.
Although \textit{Herschel} and SOFIA enable us to have an opportunity to access
such data sets for the first time, spatial coverage large enough to include
an entire Galactic star-forming region remains poor for global
understanding of PDRs, in particular for a [\ion{C}{ii}] map.

\begin{figure}[t]
 \centering
 \vspace{3mm}
 \resizebox{8cm}{!}{\includegraphics{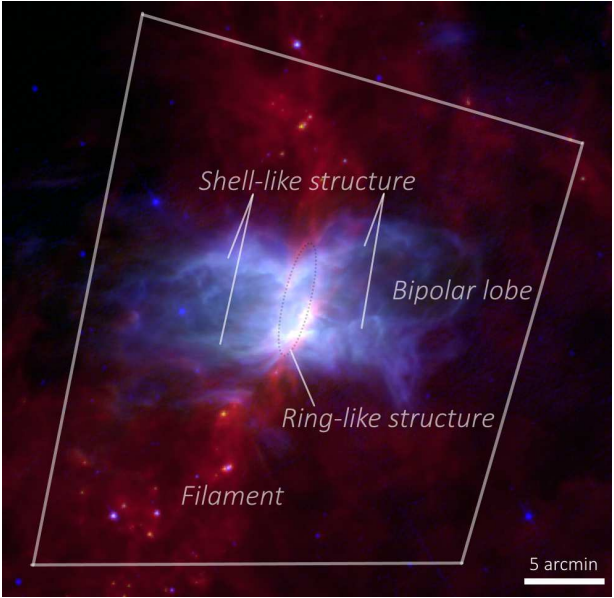}}
    \caption{[\ion{C}{ii}] observation area (solid box) of RCW~36
 superimposed on a composite image shown with the equatorial J2000
 coordinate system: AKARI~9~$\mu$m (blue), \textit{Herschel}~70~$\mu$m
 (green), and \textit{Herschel}~250~$\mu$m (red). RCW~36 is formed in a
 large-scale filament extending north-south direction. A bipolar lobe,
 shell-like structures and a ring-like structure (black dashed ellipse)
 are identified by \cite{Minier2013}.}   
    \label{fig:obsarea}
\end{figure}

RCW~36 is a Galactic massive star-forming region formed in a filament
structure of the Vela C molecular cloud complex~\citep{Hill2011}; the
\ion{H}{ii} region is formed by $\sim350$ stars including two O-type 
and one B-type stars~\citep{Ellerbroek2013}. Recent CO
observations suggest that massive stars were formed by a cloud-cloud
collision along the filament direction~\citep{Sano2018}. The composite IR
image in Fig.~\ref{fig:obsarea} shows two major features in
addition to the filament structure: 1) a bipolar lobe with shell-like
structures extending the east-west direction which is perpendicular to
the filament structure, 2) a ring-like structure (dashed ellipse) with a
major axis of 2~pc at its center. The two features are attributed to
blowout of the filament structure towards surrounding low-density
regions because of intense ionizing radiation and stellar winds from massive
stars~\citep{Minier2013}. Thus, RCW~36 showing spectacular features is a
best laboratory to understand the radiative feedback from massive stars.
In this paper, the distance to RCW~36 is taken to be
1.09~kpc~\citep{Gaia2018}.  

 \begin{figure}[t!]
   \center
  \resizebox{\hsize}{!}{\includegraphics{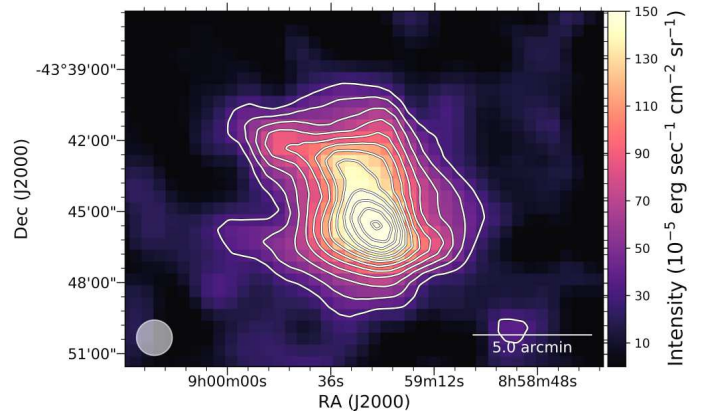}}
    \caption{[\ion{C}{ii}] intensity map of RCW~36 showing an area
 of 20\arcmin $\times$ 15\arcmin taken by FPS100. The contours are
 linearly spaced 12 levels from $3.3\times10^{-4}$ to $1.9\times10^{-3}$
 erg sec$^{-1}$ cm$^{-2}$ sr$^{-1}$. The PSF size in FWHM is
 shown in the lower left corner.}  
    \label{fig:cii} 
 \end{figure}

\section{Observations and Data analysis}

\subsection{[CII]~158$\mu$m data from T100}
We carried out [\ion{C}{ii}] observations of RCW~36 on
15 Nov.~2004, 30 Nov.~2017, and 28 Oct.~2018 by a
Fabry-P\'{e}rot spectrometer aboard a 100-cm balloon-borne far-IR
telescope with an angular resolution of
90\arcsec~\citep[hereafter FPS100:~][]{Ghosh1988, Nakagawa1998}. At the Hyderabad
Balloon Facility of the Tata Institute 
of Fundamental Research (TIFR) in India, FPS100 was launched into the
stratosphere of an altitude of $\sim$30 km. RCW~36 was then observed
with a spatially unchopped, fast spectral scan mode by two sets of the
spatial raster scans covering a large-scale area of $\sim30 \arcmin
\times 30\arcmin$ in total~(Fig.~\ref{fig:obsarea}); detailed
descriptions on FPS100 and its observation modes can be found in
\citet{Mookerjea2003}. Telescope pointing to the science target was
achieved by offsetting it with reference to a nearby bright guide star
and the absolute pointing error is estimated to be typically $\sim$ 
60\arcsec. [\ion{C}{ii}] flux calibrations were made with Orion B by
comparing its [\ion{C}{ii}] map taken by Kuiper Airborne Observatory
\citep{Jaffe1994}; the flux calibration error is $\sim10$\%.

The data reduction follows the same manner as \citet{Kaneda2013}; a
[\ion{C}{ii}] intensity map was obtained by subtracting atmospheric
background and astronomical continuum emission components from each
spectrum scan. For each set of the spatial scans taken in 2004, 2017,
and 2018, an atmospheric background spectrum was estimated by averaging the
spectral scan data at the eastern edge of the spatial scan
legs every three legs; an area of $\sim30\arcmin \times 7\arcmin$ was used as the 
background region where no significant [\ion{C}{ii}] line 
emission from RCW~36 was detected. Then, we fitted the spectral scan
data on the corresponding spatial scans by a combination of the 
atmospheric background spectrum, a Rayleigh–Jeans
regime modified black-body function with an emissivity power-law
index ($\beta$) of 2.0 for an astronomical 158~$\mu$m continuum emission, and a
Lorentzian function for the [\ion{C}{ii}] emission. Amplitudes of
the atmospheric background, the astronomical 158~$\mu$m continuum, and 
[\ion{C}{ii}] emission components were set to be free, while the
center and the width of the Lorentzian function were fixed to the values
estimated from the data of several spectral scans with high signal-to-noise
ratio. To determine the positional offset in a [\ion{C}{ii}] map, we
shifted the 158~$\mu$m continuum emission map along the R.A. and
Dec. axes so that the peak in the continuum emission map is matched
to that in a \textit{Herschel} 160~$\mu$m map. Finally, the combined
[\ion{C}{ii}] map obtained from the three maps was regridded with pixel
sizes of 90\arcsec for pixel-by-pixel 
correlation analyses. This ensures measurements for every pixel to be
independent samples (without coupling with the neighbouring pixels).

\begin{figure}[t!]
 \center
 \resizebox{\hsize}{!}{\includegraphics{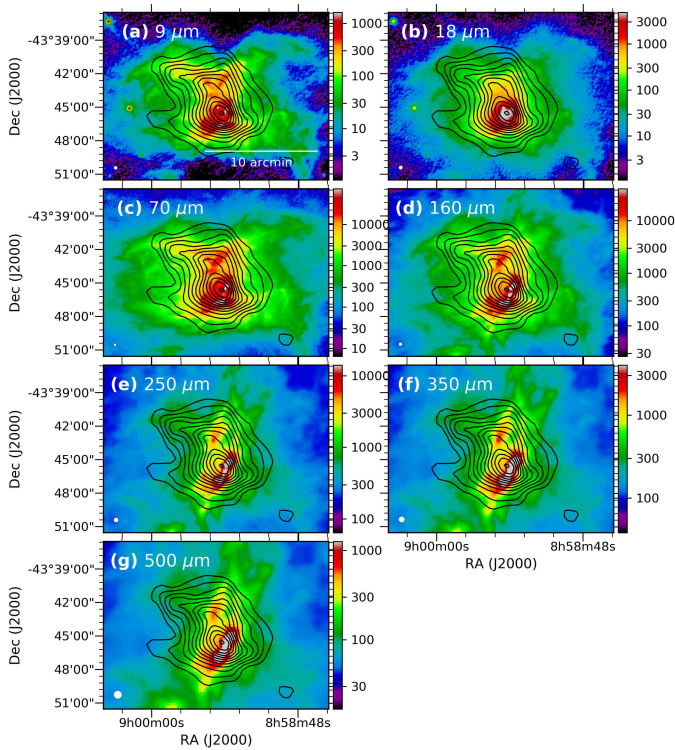}}
    \caption{Seven-band images of RCW~36 from the
 AKARI and \textit{Herschel} data in (a) AKARI~9~$\mu$m, (b)
 AKARI~18~$\mu$m, (c) \textit{Herschel}~70~$\mu$m, (d) \textit{Herschel}~160~$\mu$m, (e)
 \textit{Herschel}~250~$\mu$m, (f) \textit{Herschel}~350~$\mu$m, and (g)
 \textit{Herschel}~500~$\mu$m  bands. In each image, the color bar is given in
 units of MJy~sr$^{-1}$. The PSF size in FWHM is in the lower left corner. The
 contours superimposed on the images are the same as those in
 Fig~\ref{fig:cii}. }
    \label{fig:irim}
\end{figure}

\subsection{Broad-band IR data from AKARI and \textit{Herschel}}
RCW~36 was observed with the AKARI all-sky
survey in mid-IR wavelengths~\citep{Onaka2007, Ishihara2010} and
\textit{Herschel} as part of HOBYS (The Herschel imaging survey of OB
young stellar objects) guaranteed time key 
program~\citep{Hill2011}. For comparison with the [\ion{C}{ii}] map, we
used two-band images from AKARI~(9 and 18~$\mu$m) and five-band images
from \textit{Herschel}~(70, 160, 250, 350, and 500~$\mu$m).

The spatial resolutions of AKARI~9~$\mu$m and 18~$\mu$m are 12 and
14\arcsec, respectively. The flux calibration uncertainty of
the two bands is $\sim$10\%~(Ishihara et al. in prep.).  
The \textit{Herschel} data were taken from the \textit{Herschel} Science
Archive. The \textit{Herschel}/PACS and SPIRE data are level-2.5 products (SPG
version 14.2.0) providing JScanam maps and level-2 products (SPG version
14.1.0) providing Naive maps, respectively. The beam size has the FWHMs
of 5, 11, 18, 24, and 35\arcsec at \textit{Herschel}~70~$\mu$m, 160~$\mu$m,
250~$\mu$m, 350~$\mu$m, and 500~$\mu$m,
respectively. The systematic flux calibration uncertainty of the
\textit{Herschel} bands is applied to be $\sim$5\%~(PACS Observer's Manual
version 2.5.1 and SPIRE Observer's Manual version 2.5).

A background level for each of the seven-band IR images was estimated by
averaging pixel values in a region where no significant [\ion{C}{ii}]
emission was detected to subtract a diffuse component
that is not associated to RCW~36 from each image.
The background level is 0.6\% (AKARI 9~$\mu$m), 0.2\%
(AKARI 18~$\mu$m), 0.1\% (\textit{Herschel} 70~$\mu$m), 0.4\%
(\textit{Herschel} 160~$\mu$m), 3\% (\textit{Herschel} 250~$\mu$m), 4\%
(\textit{Herschel} 350~$\mu$m), and 5\% (\textit{Herschel} 500~$\mu$m) of  
the peak surface brightness. For multi-band
analysis, the spatial resolutions of background-subtracted IR images
were reduced to match a Gaussian PSF with the FWHM of 90\arcsec for the
FPS100 data by convolving PACS and SPIRE-band images with kernels provided
by~\cite{Aniano2011}. Because such kernels for the AKARI bands are
currently not available in the
library\footnote{http://www.astro.princeton.edu/$\sim$ganiano/Kernels.html}, 
the AKARI images were convolved by a Gaussian kernel to approximate the
FPS100 PSF. Then, the convolved images were regridded with a pixel size
of 90\arcsec matched with the PSF of the FPS100 data for
pixel-by-pixel correlation analyses.

\begin{figure}[t!]
  \vspace{-4mm}
  \centering
  \resizebox{6.5cm}{!}{\includegraphics{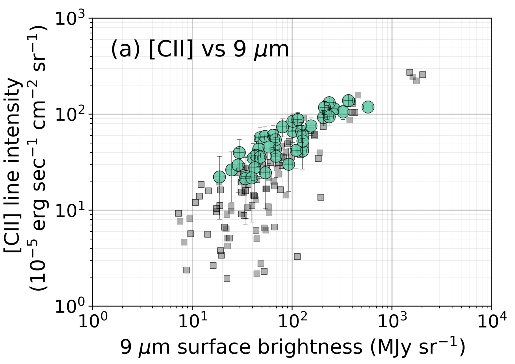}}
  \resizebox{6.5cm}{!}{\includegraphics{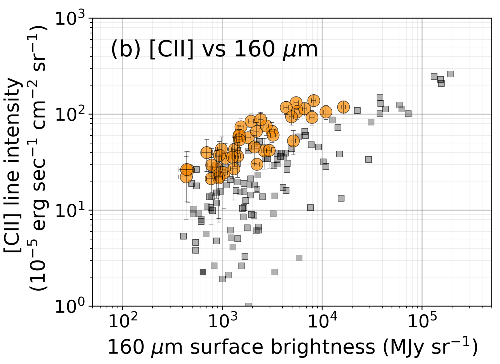}}
\caption{Correlation plots of brightness between (a) [\ion{C}{ii}] and 9~$\mu$m, and (b)
 [\ion{C}{ii}] and 160~$\mu$m for RCW~36 (circle) together with those
 for RCW~38 (square) from Fig.3 in~\citet{Kaneda2013}.}
\label{fig:correlation}
\end{figure}

\section{Results}
\subsection{[CII]~158~$\mu$m and broadband IR maps}
Figure~\ref{fig:cii} shows the [\ion{C}{ii}] intensity
map of RCW~36 by FPS100. The 1-sigma fluctuation per pixel is
$1.1\times10^{-4}$ erg~sec$^{-1}$~cm$^{-2}$~sr$^{-1}$. The
[\ion{C}{ii}] emission overall extends toward the east-west direction from the
peak, which is almost perpendicular to the direction of the filament
structure as shown in Fig.~\ref{fig:obsarea}. 

The seven-band IR images by AKARI and \textit{Herschel} are shown in
Fig.~\ref{fig:irim}. The AKARI~9~$\mu$m image mainly traces emission from
PAHs and shows a bipolar lobe with 
shell-like structures in the east-west direction~(Fig.~\ref{fig:obsarea}). The overall
spatial distribution of the AKARI~9~$\mu$m map is in good agreement
with that of the \textit{Herschel}~70--500~$\mu$m maps which trace emission
from large grains, although the \textit{Herschel}~250--500~$\mu$m maps
show more prominent filament structure in addition to
the bipolar lobe. In the AKARI~18~$\mu$m image which traces emissions
from large and very small grains, the spatial distribution also shows
the bipolar lobe but does not show shell-like structures clearly.

\begin{figure}[t!]
 \center
 \resizebox{\hsize}{!}{\includegraphics{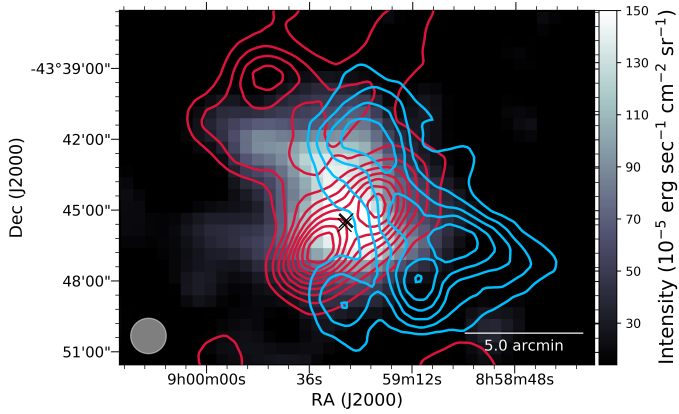}}
    \caption{Same as Fig.~\ref{fig:cii}, but $^{12}$CO($J$=2--1)
 contours~\citep[Fig.~4a in][]{Sano2018} and positions of OB stars
 (cross)~\citep{Ellerbroek2013}. The cyan and red colored contours
 represent two-velocity molecular cloud components with
 $V_\mathrm{LSR}=$4.1--6.1~km~s$^{-1}$ (blue cloud) and
 7.6--12.0~km~s$^{-1}$ (red cloud), respectively. The spatial
 resolution of the $^{12}$CO($J$=2--1) map taken by NANTEN2 is similar
 to the [\ion{C}{II}] map~(90\arcsec).} \label{fig:ciico}
\end{figure}

To compare spatial distributions between the [\ion{C}{ii}] map and a
broadband IR emission map, contours of the [\ion{C}{ii}] intensity are
superimposed on each IR map~(Fig.~\ref{fig:irim}). The [\ion{C}{ii}] map
shows two prominent arms extending from its peak to the north-east and
south-west directions, which are spatially in good agreement with
shell-like structures of the bipolar lobe observed in the broadband IR images. 
The peak position in the [\ion{C}{ii}] map is located close to the IR bright
rims along the ring-like structure~\citep{Minier2013}. For quantitative
comparison of spatial distributions, Fig.~\ref{fig:correlation} shows pixel-by-pixel
correlation plots of (a) [\ion{C}{ii}] vs. 9~$\mu$m and (b)
[\ion{C}{ii}] vs. 160~$\mu$m. Here, we removed pixels with
the optical depth at $\lambda=9$~$\mu$m higher than unity from
correlation plots. Since the AKARI 9~$\mu$m 
band is overlapping with the interstellar silicate feature around
9.7~$\mu$m, the interstellar extinction in the AKARI 9~$\mu$m band is
the most severe among the seven IR bands. To evaluate the extinction,
the optical depth at $\lambda=9$~$\mu$m, $\tau_{9}$ is calculated from
$\tau_{9}=4.8\times10^{-2}A_\mathrm{V}$ by applying
$A_9/A_\mathrm{Ks}=5.8\times10^{-1}$~\citep{Xue2016} and
$A_\mathrm{Ks}/A_\mathrm{V}=8.9\times10^{-2}$~\citep{Glass1999};
$\tau_9>1$ corresponds to $A_\mathrm{V}>21$~mag. \cite{Hill2011}
investigated an $N_\mathrm{H_2}$ column density map of the Vela C
molecular complex including RCW~36 based on pixel-by-pixel SED
fitting in the wavelength range of 70--500~$\mu$m. We then converted the
$N_\mathrm{H_2}$ column density to visual extinction units assuming
$N_\mathrm{H_2} = 0.94\times10^{21} A_\mathrm{V}$
cm$^{-2}$~\citep{Bohlin1978}. From the map around RCW~36, the regions
with $A_\mathrm{V}>21$~mag are located around the peak seen in the
\textit{Herschel} 500~$\mu$m map. Therefore, the relevant pixels showing
$\tau_9>1$ were removed.  
%
%
%
%
Clearly, the [\ion{C}{ii}]--9~$\mu$m relation for RCW~36 is in good agreement
with that for RCW~38 as denoted by black-filled squares, while the
[\ion{C}{ii}]--160~$\mu$m relation for RCW~36 shows higher brightness
ratio of [\ion{C}{ii}]/160~$\mu$m than that for RCW~38.

\begin{figure}[t!]
 \vspace{0mm}
 \resizebox{\hsize}{!}{\includegraphics{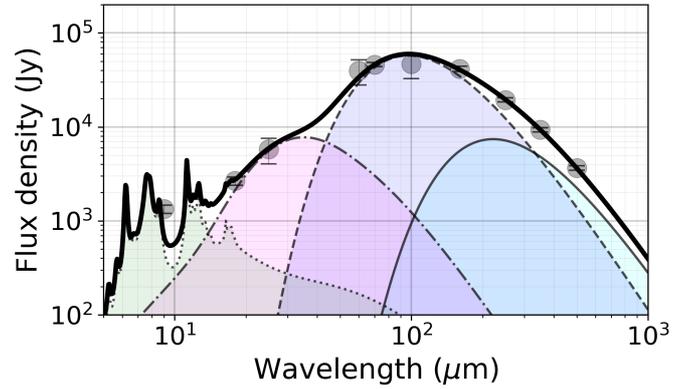}}
 \caption{Global SED of the whole region of RCW~36 obtained from the aperture radius
 of 25\arcmin~at the center position of RA= 8:59:30.963,~Dec= -43:46:56.270 (J2000.0). To obtain the eight-free parameters, the
 SED fitting was performed for the original seven-band data plus IRAS~25~$\mu$m,
 60~$\mu$m, and 100~$\mu$m data. The thick solid line
 shows the best-fitting model, which is described in Eq.~(1). The very
 cold dust, cold dust, warm dust, and PAH components are denoted by thin
 solid, dashed, dash-dotted, and dotted lines, respectively. } 
\label{fig:sed}
 \vspace{-4mm}
\end{figure}

\subsection{[CII]~158~$\mu$m and molecular gas maps}
\cite{Sano2018} presented the latest CO observations of RCW~36 and found
two molecular clouds at the velocities
$V_\mathrm{LSR}\sim5.5$~$\mathrm{km\ s^{-1}}$~(blue cloud) and
9~$\mathrm{km\ s^{-1}}$~(red cloud), which are likely to be 
physically associated with RCW~36. They also showed
that there is no bridge-like feature connecting the two molecular clouds
in velocity space. Figure~\ref{fig:ciico} shows $^{12}\mathrm{CO}$($J$=2--1)
contours~\citep[Fig.~4a in][]{Sano2018} overlaid on the [\ion{C}{II}]
image. The [\ion{C}{II}] emission is not spatially in good agreement
with the $^{12}\mathrm{CO}$ emission for both red and blue cloud
components; the bright $^{12}\mathrm{CO}$ emission from the dense red
cloud component has the double peak and its structure is elongated along
the filament structure. The double peak is located around the
[\ion{C}{II}] peak and the massive stars as denoted by the cross
marks. \cite{Sano2018} found that the dense red cloud coincides with the 
star cluster, while the diffuse blue cloud does not. The red cloud seems
to be physically close to the star cluster and its proximate surface
ionized by the UV radiation from these stars, while the blue cloud is
less affected by UV from this cluster due to its far location. 
Moreover, the diffuse CO emission extending from the filament structure
to eastern (red cloud) 
and western (blue cloud) directions tends to be distributed in the outer
rim of shell-like structures seen in [\ion{C}{II}] and broadband IR
maps. 

From the above situation in spatial distributions, we propose a possible
geometry of the [\ion{C}{II}] emission region with respect to the two
collided molecular clouds as follows: for the
positional relation among the observer, the red cloud, and the blue
cloud, there are two possible cases. The first case is that the red cloud
is closer to the observer when the red and blue clouds are colliding with
each other, while the second case is that the blue cloud is closer to the
observer when the red cloud went through the blue one after collision. As
shown in \cite{Fukui2016}, colliding two clouds are generally connected
to each other in velocity space and show a bridge-like feature in a
position-velocity diagram. Because the bridge-like feature was not
observed in RCW36~\citep{Sano2018}, the observational result supports
the second case. Therefore, the dense red cloud with the molecular
filament went through the blue cloud, and is then associated with the
massive stars formed by the cloud-cloud collision, while the 
diffuse red and blue clouds are mostly located the outer rim of
shell-like structures. The massive stars formed by the collision between
the red and blue clouds illuminate surface of those two clouds. Given
that the most of the [\ion{C}{ii}] emission originates from those far-UV
illuminated surface of clouds, the [\ion{C}{ii}]-emitting region is
expected to be facing toward observers without significant attenuation
by foreground dust grains associated with CO clouds.




\begin{figure}[t!]
 \resizebox{\hsize}{!}{\includegraphics{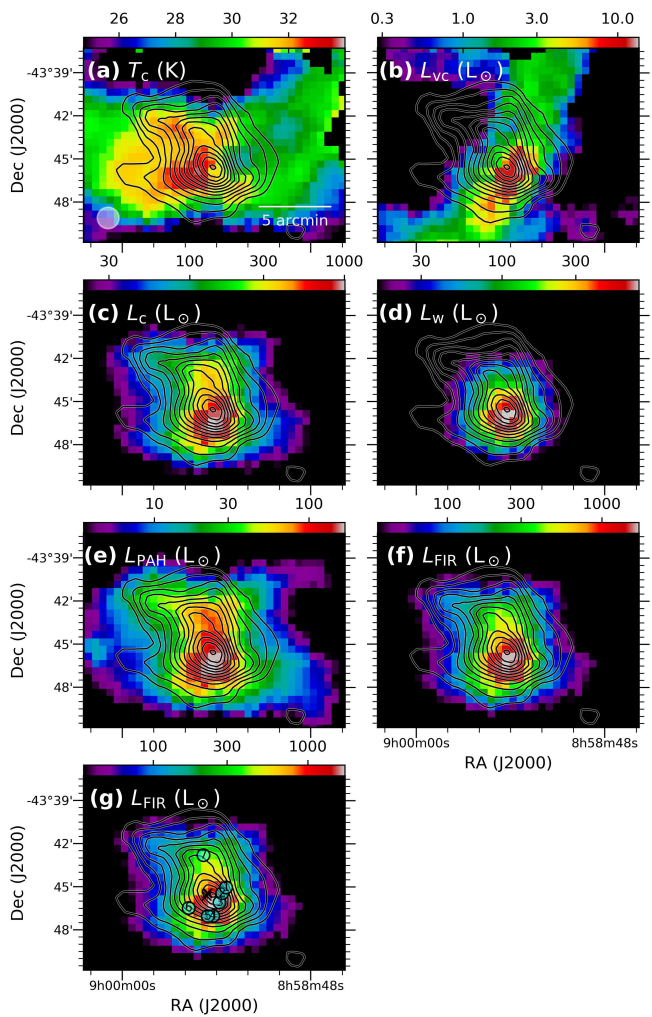}}
 \caption{Spatial distributions of (a) cold dust temperature
 ($T_\mathrm{c}$), (b) very cold dust luminosity ($L_\mathrm{vc}$), (c)
 cold dust luminosity ($L_\mathrm{c}$), (d) warm dust luminosity
 ($L_\mathrm{w}$), (e) PAH luminosity ($L_\mathrm{PAH}$), and (f) far-IR
 luminosity
 ($L_\mathrm{FIR}$=$L_\mathrm{vc}+L_\mathrm{c}+L_\mathrm{w}$). The panel
 (g) is same as the panel (f), but the positions of seven clumps and OB stars are denoted 
 by circles and crosses, respectively~\citep{Ellerbroek2013, Minier2013}. 
 Dust temperature and luminosity are in units of kelvin and solar
 luminosity, respectively. 
 For each luminosity map, the color scale ranges from 2\% to 98\% of the
 peak luminosity. The contours superimposed on the images are the same
 as those in Fig.~\ref{fig:cii}. 
}
 \label{fig:sedfitres}
\end{figure}

\section{Discussion}
  \subsection{Spectral decomposition into very cold dust, cold dust,
  warm dust and PAH components}  
  Spectral decomposition analysis on a pixel-by-pixel basis enables us to
  investigate spatial distributions of PAH and dust properties and
  the relation between those properties and [\ion{C}{ii}] emission. An
  individual SED constructed from the seven-band fluxes at each pixel
  (pixel size of 90\arcsec) is reproduced by a triple-component modified
  blackbody plus a PAH model expressed as   
\begin{equation}
 \begin{split}
 F_{\nu,\mathrm{IR}} &= A_\mathrm{PAH}F_{\nu,\mathrm{PAH}} +
  A_\mathrm{vc} \nu^{\beta_\mathrm{vc}} B_{\nu}(T_\mathrm{vc}) +
  A_\mathrm{c} \nu^{\beta_\mathrm{c}} B_{\nu}(T_\mathrm{c}) \\
  &\quad + A_\mathrm{w} \nu^{\beta_\mathrm{w}}
 B_{\nu}(T_\mathrm{w})\left\{1+
  \left(\frac{\nu_c}{\nu}\right)^3\nu^{-\alpha}
  e^{-\left(\frac{\nu_c}{\nu}\right)^2} \right\},  
\label{eq1}
 \end{split}
\end{equation}
where $T$, $\beta$, $A$, and $B_\nu(T)$ are the dust temperature, the dust
emissivity power-law index, amplitude, and the Planck function,
respectively. Suffixes vc, c, w, and PAH denote very cold dust, cold
dust, warm dust and PAH components, respectively. The $\beta$ value for
each dust component is assumed to be 2.0 as a typical value of the
Galactic ISM~\citep{Anderson2012}. The second term in the
warm dust component is the analytic approximation of
thermal emission from dust grains which are exposed to a
range of starlight intensities: dust emissions with
different dust temperatures assuming a power-law temperature
distribution to take the hot dust component into account in mid-IR
wavelengths~\citep{Casey2012}. The power-law turnover frequency $\nu_c$
defined by~\cite{Casey2012} is a function of the mid-IR power-law slope
$\alpha$ and $T_\mathrm{w}$. The flux density of the PAH component,
$F_\mathrm{PAH}(\nu)$, is calculated as described in \cite{suzuki2010}
and is based on the PAH parameters taken from \cite{li} and \cite{li2}
by assuming the PAH size distribution ranging from 3.55 to 300 \AA, the
fractional ionization and the temperature probability distribution for
the typical diffuse ISM with the interstellar radiation field in the
solar neighborhood. PAHs with sizes larger than 15~\AA~contribute to
$\sim20$~$\mu$m continuum emission~\citep{li2}. Since very small grains
(VSGs) which are stochastically heated by absorbed far-UV photons contribute to
$\sim20$~$\mu$m continuum emission, the PAH component in Eq.~(\ref{eq1})
takes the VSG emission into account.


\begin{figure*}[t!]
\centering
\includegraphics[width=6.5cm]{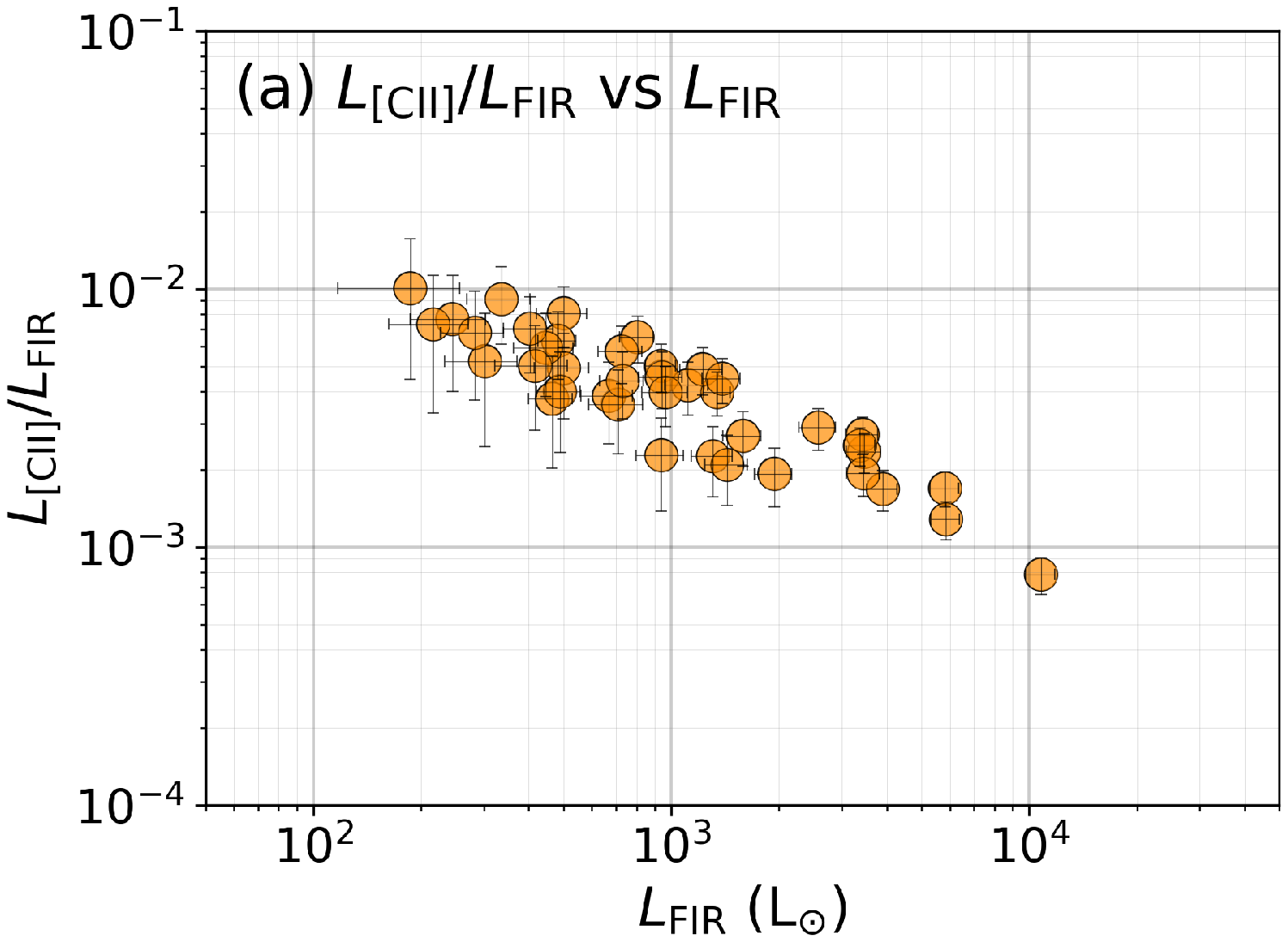}
\includegraphics[width=6.5cm]{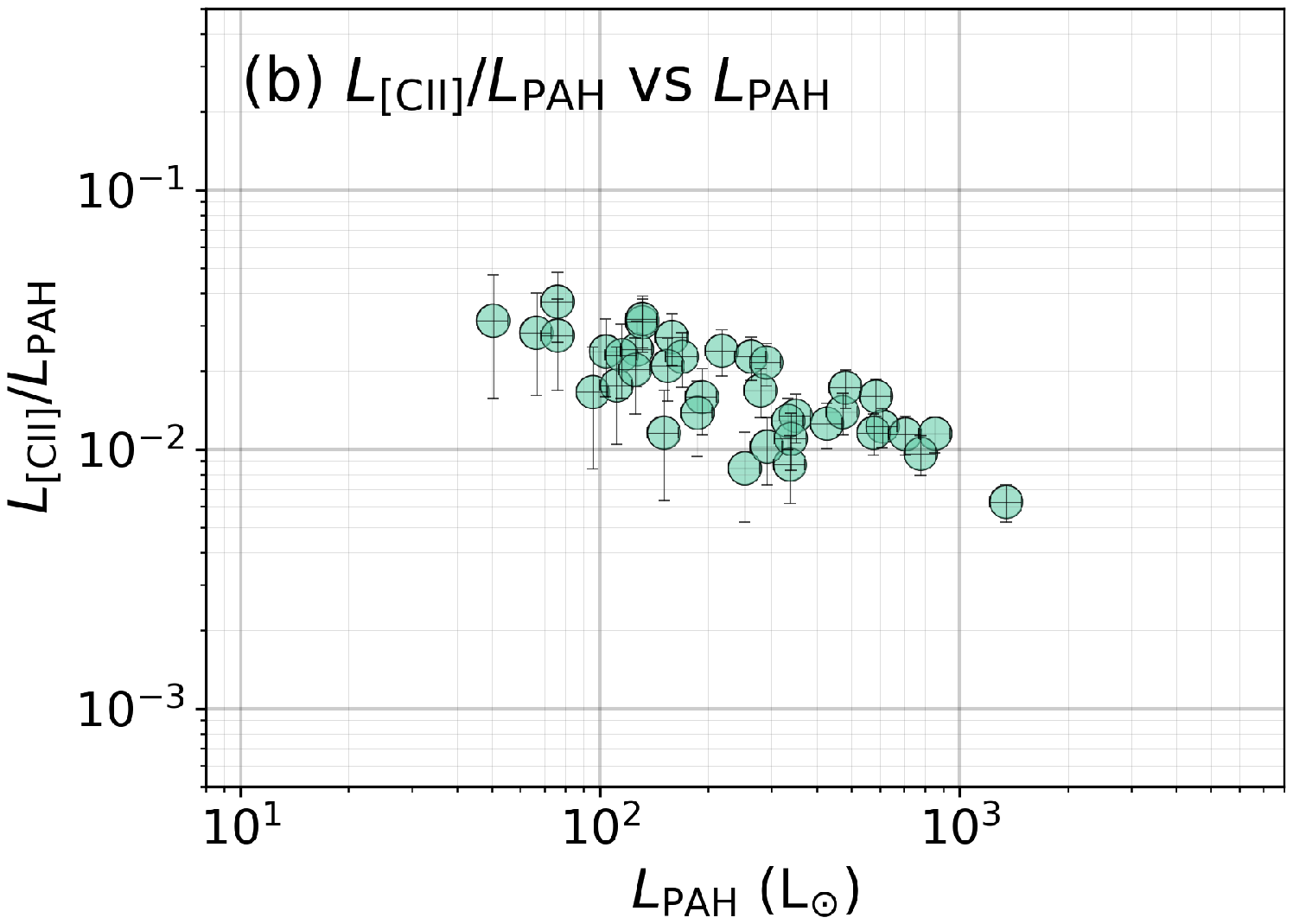}
 \caption{Correlation plots of (a) $L_\mathrm{[CII]}/L_\mathrm{FIR}$ vs. $L_\mathrm{FIR}$ and
 (b) $L_\mathrm{[CII]}/L_\mathrm{PAH}$ vs. $L_\mathrm{PAH}$.}
\label{fig:lum_correlation}
\end{figure*}

The eight-free parameters for the SED model cannot be determined from a
seven-band data set. We then fixed $T_\mathrm{vc}$ and 
$\alpha$ values obtained from a global SED of RCW~36 by adding IRAS~25~$\mu$m,
60~$\mu$m, and 100~$\mu$m data. Since the parameters
$T_\mathrm{w}$ and $A_\mathrm{w}$ are both determined by the 18~$\mu$m
flux density, they are fully degenerate in the model. To avoid this
degeneracy in the calculation of the spatial distribution of luminosity
for the warm dust component, we adopted a fixed $T_\mathrm{w}$ value
obtained from the global SED fitting.
Fig.~\ref{fig:sed} shows the global SED together with the best-fit
model. The best-fit parameters of $T_\mathrm{vc}$, $T_\mathrm{c}$,
$T_\mathrm{w}$, and $\alpha$ are 13.1~K, 29.9~K, 77.9~K, and 2.9,
respectively. The obtained $T_\mathrm{vc}$ is consistent with the result
measured at dense filaments around RCW~36~\citep{Hill2011}. 
Initial values of pixel-by-pixel SED fitting were applied with the best-fit
values obtained from fitting for the global SED. Then, an SED fitting
procedure was performed to minimize $\chi^2$ at each pixel.  As a
result, the five-free ($T_\mathrm{c}$, $A_\mathrm{vc}$, $A_\mathrm{c}$,
$A_\mathrm{w}$, and $A_\mathrm{PAH}$) and the three-fixed
($T_\mathrm{w}$, $T_\mathrm{vc}$, and $\alpha$) parameters better
reproduce the observed fluxes with pixel-averaged $\chi^2_\nu$~(two
degrees of freedom, DOF) of 0.4. Resultant pixel-averaged errors on
fitted parameters are 3\% ($T_\mathrm{c}$), 75\% ($A_\mathrm{vc}$), 17\%
($A_\mathrm{c}$), 16\% ($A_\mathrm{w}$), and 12\% ($A_\mathrm{PAH}$).      

Figure~\ref{fig:sedfitres} shows spatial distributions of cold dust
temperature ($T_\mathrm{c}$), very cold dust, cold
dust, warm dust, and PAH luminosities ($L_\mathrm{vc}$, $L_\mathrm{c}$,
$L_\mathrm{w}$, and $L_\mathrm{PAH}$) integrated between 3--1000~$\mu$m
from the best-fit SED model, and the far-IR ($L_\mathrm{FIR}$) luminsity
is calculated from $L_\mathrm{vc}+L_\mathrm{c}+L_\mathrm{w}$.
In the case of pixel-by-pixel SED fitting based on the seven-band IR images without the
background subtraction, the obtained luminosities are systematically increased by
$\sim1\%$ along the filament and by $\sim10\%$ outside the filament, 
except for $L_\mathrm{vc}$: $\sim50\%$ and $\sim500\%$ along and outside
the filament, respectively. Because the contribution of
$L_\mathrm{vc}$ to the total luminosity is significantly smaller than every other
components, a factor five uncertainty in $L_\mathrm{vc}$ gives a
negligible impact on the inferred physical parameters. In the following discussions, the
key physical parameters are $L_\mathrm{[\ion{C}{ii}]}/L_\mathrm{FIR}$ and 
$L_\mathrm{[\ion{C}{ii}]}/L_\mathrm{PAH}$. Since we will discuss an
order-of-magnitude variation in those key parameters along and outside
the filament regions, the systematic change does not affect our
conclusions. Therefore, the luminosities obtained from the background
subtraction are applied in the following discussions.
The spatial distribution of
$T_\mathrm{c}$, which is in good agreement with 
that shown in \citet{Hill2011}, shows higher $T_\mathrm{c}$ values along
the shell-like structures of the bipolar lobe and the ring-like
structure caused by the stellar winds from the central OB
stars~\citep{Minier2013}, while the map shows lower $T_\mathrm{c}$ along 
the filament structure in the north-south direction. 

As for luminosity maps, the spatial distribution of $L_\mathrm{PAH}$ is
very similar to that of $L_\mathrm{c}$. The [\ion{C}{II}] emission is
spatially correlated with both $L_\mathrm{PAH}$ and $L_\mathrm{c}$, but
is relatively poorly correlated with $L_\mathrm{w}$ and not for
$L_\mathrm{vc}$. Overall spatial distributions of $L_\mathrm{PAH}$ and
$L_\mathrm{c}$ show clear shell structures of the bipolar lobe and the
ring-like structure, while that of $L_\mathrm{w}$ is concentrated around
the position of the central OB stars~\citep{Ellerbroek2013}. 
To quantify the spatial correlation, a
multiple linear regression analysis was performed using the Python
StatsModels package~\citep{seabold2010}. Since $L_\mathrm{PAH}$ and
$L_\mathrm{c}$ are well correlated with each other (correlation
coefficient of 0.97), both parameters were not included as explanatory
variables for the analysis. To make the variance infraction factor (VIF)
small, we chose [\ion{C}{ii}] luminosity, $L_\mathrm{[\ion{C}{II}]}$ as
the dependent variable, and $L_\mathrm{PAH}$ and $L_\mathrm{w}$ as
explanatory variables. As a result, the adjusted coefficient of determination
$R^2$ is obtained to be 0.85 with DOF=39 and VIF=3.5, and the
correlation to the $L_\mathrm{PAH}$ variable is significant ($p<0.001$),
while that to the $L_\mathrm{w}$ variable is not ($p=0.47$) based on the
$t$-test with the 95\% confidence level. Given that the PAH emission
dominantly comes from PDRs, a good spatial correlation to
$L_\mathrm{PAH}$ suggests that the [\ion{C}{II}] emission dominantly
comes from PDRs. The good spatial correlation seen with $L_\mathrm{c}$
is also consistent, since the cold dust temperature of $\sim30$~K is
reasonable for a PDR. However, a poor spatial correlation to
$L_\mathrm{w}$ indicates that the warm dust component mostly trace
\ion{H}{II} regions surrounding the OB stars.
For the very cold dust component, it is spatially in good
agreement with the dense filament structure. The $L_\mathrm{vc}$
luminous region around the [CII] emission peak corresponds to the
positions of the $^{12}$CO peaks and massive clumps identified by
\cite{Minier2013}. Thus, the very cold dust component well traces
dense molecular regions.

\subsection{Variations of
  $L_\mathrm{[\ion{C}{ii}]}/L_\mathrm{FIR}$ and $L_\mathrm{[\ion{C}{ii}]}/L_\mathrm{PAH}$}
\begin{figure}[t!]
  \vspace{-3mm}
 \centering
 \resizebox{\hsize}{!}{\includegraphics[width=8.5cm]{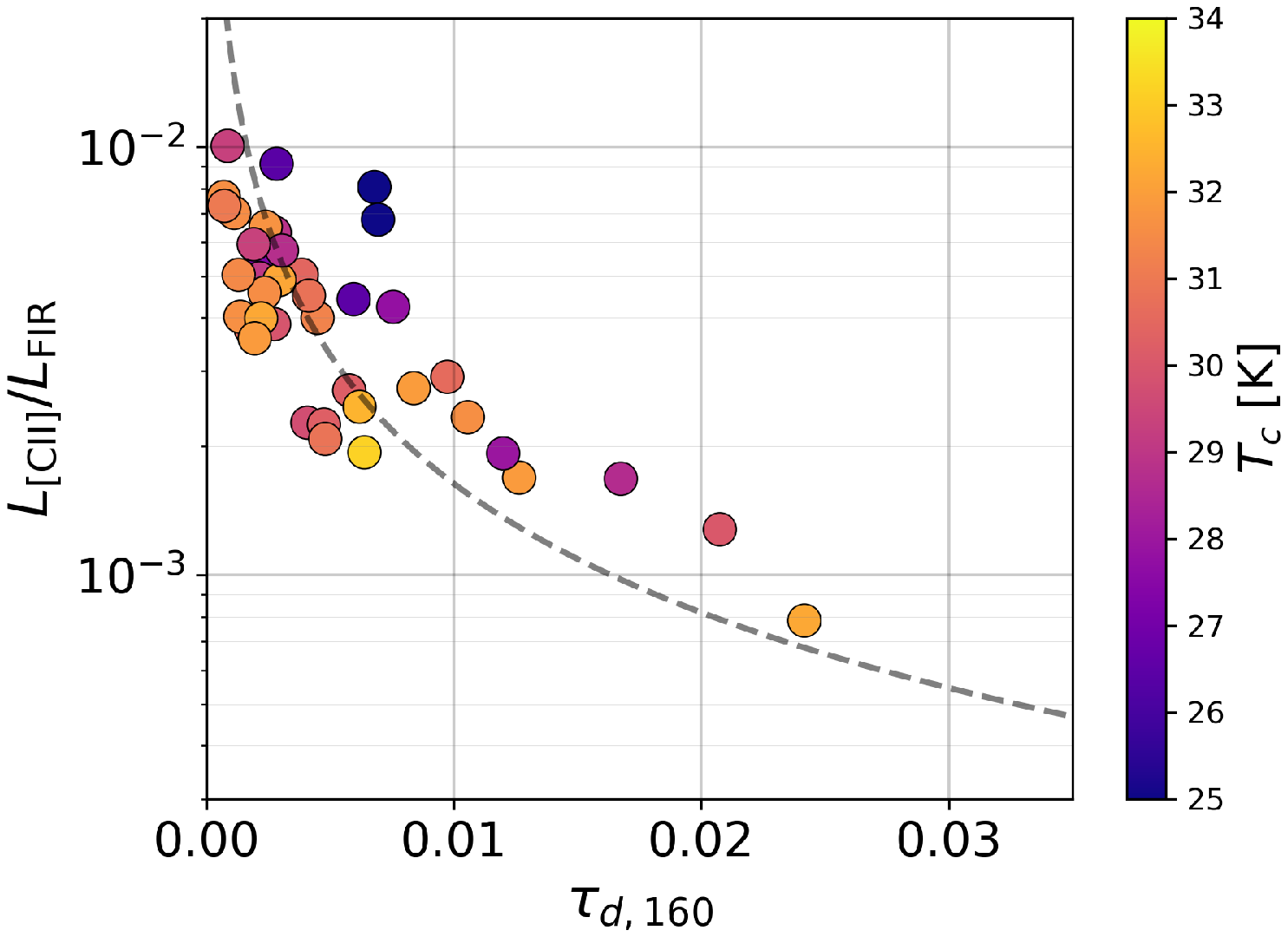}}
\caption{Correlation plot of $L_\mathrm{[\ion{C}{ii}]}/L_\mathrm{FIR}$
 vs. $\tau_\mathrm{d,160}$. The dashed line shows the relation with the
 model of a simple face-on slab of dust with the [\ion{C}{ii}] foreground
 emission~\citep{Goicoechea2015}.} 
 \label{fig:LciidLfir_vs_tau}
\end{figure}

In Fig.~\ref{fig:lum_correlation},
$L_\mathrm{[\ion{C}{ii}]}/L_\mathrm{FIR}$ varies about one order of
magnitude, while $L_\mathrm{[\ion{C}{ii}]}/L_\mathrm{PAH}$ shows a
smaller variation. Such large variation in
$L_\mathrm{[\ion{C}{ii}]}/L_\mathrm{FIR}$ is also observed in the Milky
Way and nearby
galaxies~\citep[e.g.][]{Wright1991,Stacey1991,Malhotra2001,Croxall2012,
Kramer2013, Goicoechea2015, Smith2017}. According to
\cite{Goicoechea2015}, the large variation observed toward the Orion
molecular cloud 1 (OMC1) can be explained by those in the total dust-cloud
column relative to the [\ion{C}{ii}]-emitting column along each line of
sight (geometry effect). To verify whether such
geometry effects can also cause a large range of
$L_\mathrm{[\ion{C}{ii}]}/L_\mathrm{FIR}$ in RCW~36, the dust opacity close to the [\ion{C}{ii}] 158~$\mu$m line is calculated as
$\tau_\mathrm{d,160}=10^{-26}A_\mathrm{c}\nu_{160}^{\beta_c}\Omega_\mathrm{pix}^{-1}$
for an optically-thin condition, where $\Omega_\mathrm{pix}$ is the
solid angle subtended by each
pixel~(90\arcsec). Figure~\ref{fig:LciidLfir_vs_tau} shows
$L_\mathrm{[\ion{C}{ii}]}/L_\mathrm{FIR}$ as a function of
$\tau_\mathrm{d,160}$, color-coded according to $T_\mathrm{c}$. The
dashed line corresponds to the relation calculated from a simple face-on
slab model assuming a uniform dust temperature:
$L_\mathrm{[\ion{C}{ii}]}/L_\mathrm{FIR}=
C(1-e^{-\tau_\mathrm{d,160}})^{-1}\simeq C\tau_\mathrm{d,160}^{-1}$~\citep[see Fig.~16a
in][]{Goicoechea2015}, where $C$ is a constant determined so that the
model curve intercepts the median
$L_\mathrm{[\ion{C}{ii}]}/L_\mathrm{FIR}$ and $\tau_\mathrm{d,160}$ values.
Clearly, the measured data are distributed along the model line because
of small variation in the cold dust temperature.
Therefore, large variations in $L_\mathrm{[\ion{C}{ii}]}/L_\mathrm{FIR}$
toward RCW~36 are explained by the geometry effect like the case
in OMC1.

\begin{figure}[t!]
 \centering
 \resizebox{\hsize}{!}{\includegraphics[width=8.5cm]{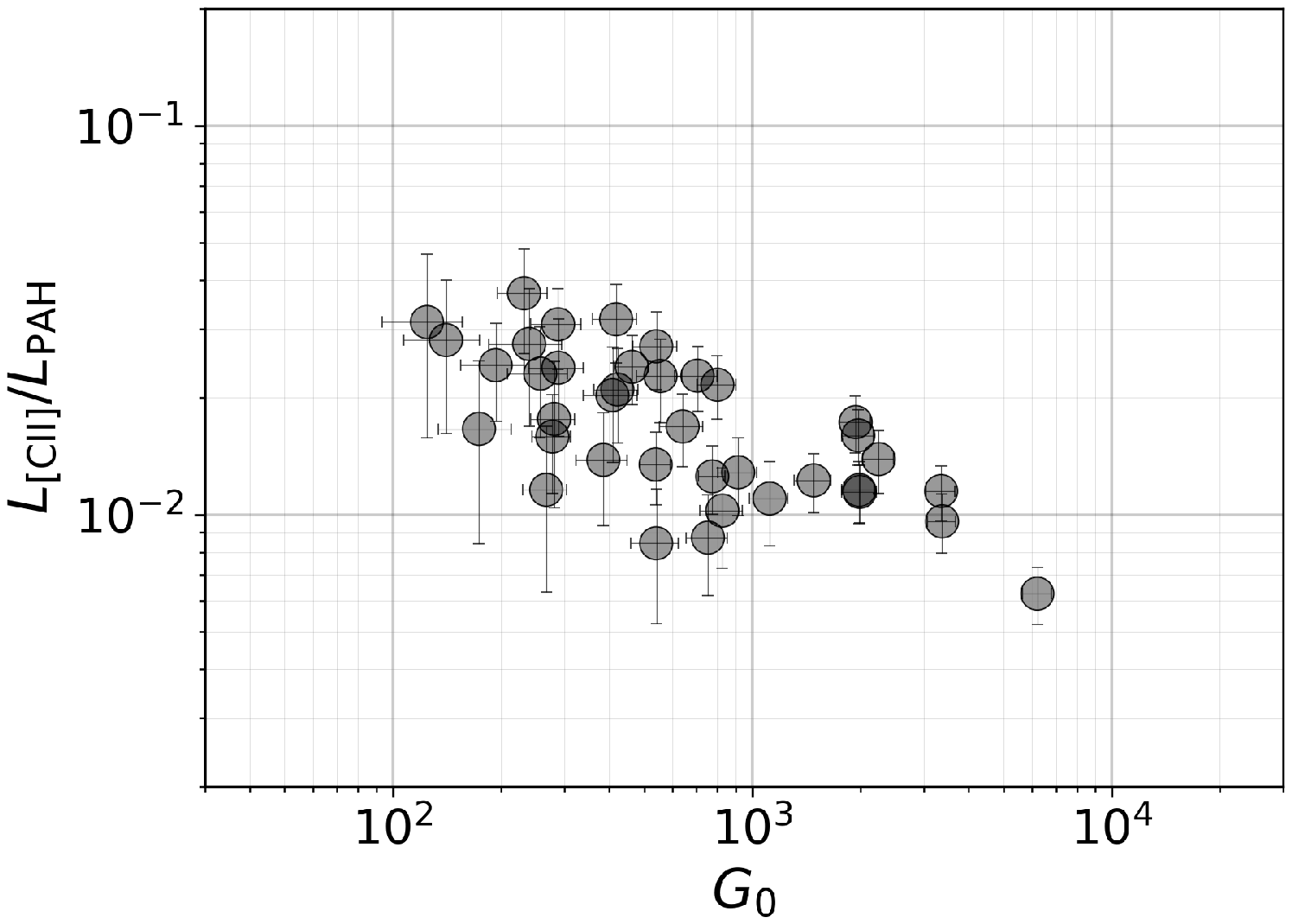}}
\caption{Correlation plot of $L_\mathrm{[\ion{C}{ii}]}/L_\mathrm{PAH}$ vs. $G_0$}  
 \label{fig:LciidLpah_vs_G0}
\end{figure}

From the spatial correlation analysis among [\ion{C}{ii}],
PAH, warm, cold and very cold dust emissions, [\ion{C}{ii}] and PAH
emissions are spatially in good agreement with each other and thus are
considered to mainly arise from the far-UV illuminated
face of clouds. Then, significant variation in 
$L_\mathrm{[\ion{C}{ii}]}/L_\mathrm{PAH}$ is not expected from the
geometry effect. Thus, another factor seems to affect  
the observed $L_\mathrm{[\ion{C}{ii}]}/L_\mathrm{PAH}$ variation. In such
far-UV dominated regions (PDRs), PAHs are expected to play a significant
role of photo-electric heating of gas~\citep{Bakes1994}. However, its heating
efficiency is decreased as the far-UV radiation field increases due to
the presence of larger amount of ionized
PAHs~\citep{Okada2013}. To verify if photo-electric
heating efficiency can explain the observed variation of
$L_\mathrm{[\ion{C}{ii}]}/L_\mathrm{PAH}$, we calculated the intensity
of the incident far-UV radiation field $G_0$ in Habing units by using
the following equation~\citep{Hollenbach1999}: $G_0\sim
I_\mathrm{FIR}/2.6\times10^{-4}~\mathrm{erg\ sec^{-1}\ 
cm^{-2}\ sr^{-1}}$, where $I_\mathrm{FIR}$ is the far-IR intensity
obtained from $L_\mathrm{FIR}$ at each pixel. $G_0$ values thus obtained
show the range of $\sim10^2$--$10^4$; the peak $G_0$ in
RCW36 is $\sim6$ times lower than that in RCW~38~\citep{Kaneda2013}.
As expected, $L_\mathrm{[\ion{C}{ii}]}/L_\mathrm{PAH}$ decreases as
$G_0$ increases as shown in
Fig.~\ref{fig:LciidLpah_vs_G0}. Based on this measured
variation of $G_0$ across RCW~36, we explored the range of values of
charging parameter which is relevant for photo-electric heating
efficiency. We estimated the charging parameter $\gamma =
G_0\sqrt{T_\mathrm{g}}/n_\mathrm{e}$, where $T_\mathrm{g}$ and
$n_\mathrm{e}$ are the gas temperature and the electron density in
PDRs. The electron density is calculated from the following equation with
assumptions that (a) the electrons in PDRs are all provided by ionized
carbons and the carbon atoms are fully ionized~\citep[$n_\mathrm{e} =
1.6\times10^{-4} n_\mathrm{H}$,][]{Sofia2004} and (b) pressure between \ion{H}{ii}
regions and PDRs is balanced~($2
n_\mathrm{e,\ion{H}{ii}}T_\mathrm{e,\ion{H}{ii}} \simeq n_\mathrm{H} 
T_\mathrm{g}$); 
\begin{equation}
  \begin{split}
   n_\mathrm{e} =
   3.2\times10^{-4}n_\mathrm{e,\ion{H}{ii}}\left( \frac{T_\mathrm{e,\ion{H}{ii}}}{T_\mathrm{g}}\right), 
   \label{eq3}
  \end{split}
\end{equation}
where $n_\mathrm{H}$, $n_\mathrm{e,\ion{H}{ii}}$, and $T_\mathrm{e,\ion{H}{ii}}$
are hydrogen gas density in PDRs, electron density, and electron
temperature in \ion{H}{ii} regions, respectively. Moreover,
$n_\mathrm{e,\ion{H}{ii}}(G_0)$ is derived by solving $G_0$ at
a distance equal to the Str\"omgren sphere radius and by assuming the
fraction of luminosity above 6~eV equal to unity~\citep{Tielens2005} as 
\begin{equation}
  \begin{split}
   n_\mathrm{e,\ion{H}{ii}} (G_0) \simeq
   0.78G_0^{\frac{3}{4}}\left(\frac{5\times10^{49}~\mathrm{photons~sec^{-1}}}{N_\mathrm{Lyc}}\right)^{-\frac{1}{2}}\\\times\left(\frac{L_\star}{7.6\times10^5~\mathrm{L_\sun}}\right)^{-\frac{3}{4}},
   \label{eq4}
  \end{split}
\end{equation}
where $N_\mathrm{Lyc}$ and $L_\star$ are the total number of
ionizing photons by a star and the stellar luminosity,
respectively. When we apply $N_\mathrm{Lyc}=10^{48}$~photons sec$^{-1}$
for the spectral type of O9 as the dominant exciting star~\citep{Binder2018}, 
$L_\star=8\times10^4$~$L_\sun$~\citep{Thompson1984},
$T_\mathrm{e,\ion{H}{ii}}=7600$~K~\citep{Shaver1970}, and measured $G_0$
values, $\gamma$ ranges from $\sim10^3$ to $10^5$ $\sqrt{\mathrm{K}}\
\mathrm{cm}^{-3}$ for typical $T_\mathrm{g}$ of 10$^2$--10$^3$~K in
PDRs~\citep{Kaufman1999}. The obtained $\gamma$ range meets the
transition where the charge state of PAHs changes from neutral to fully
ionized states~\citep{Okada2013}. Therefore, the
$L_\mathrm{[\ion{C}{ii}]}/L_\mathrm{PAH}$ variation can be explained by
the variation of the photo-electric heating efficiency on PAHs.

\begin{figure*}[t!]
\centering
\hspace{-15mm}
 \includegraphics[width=14cm]{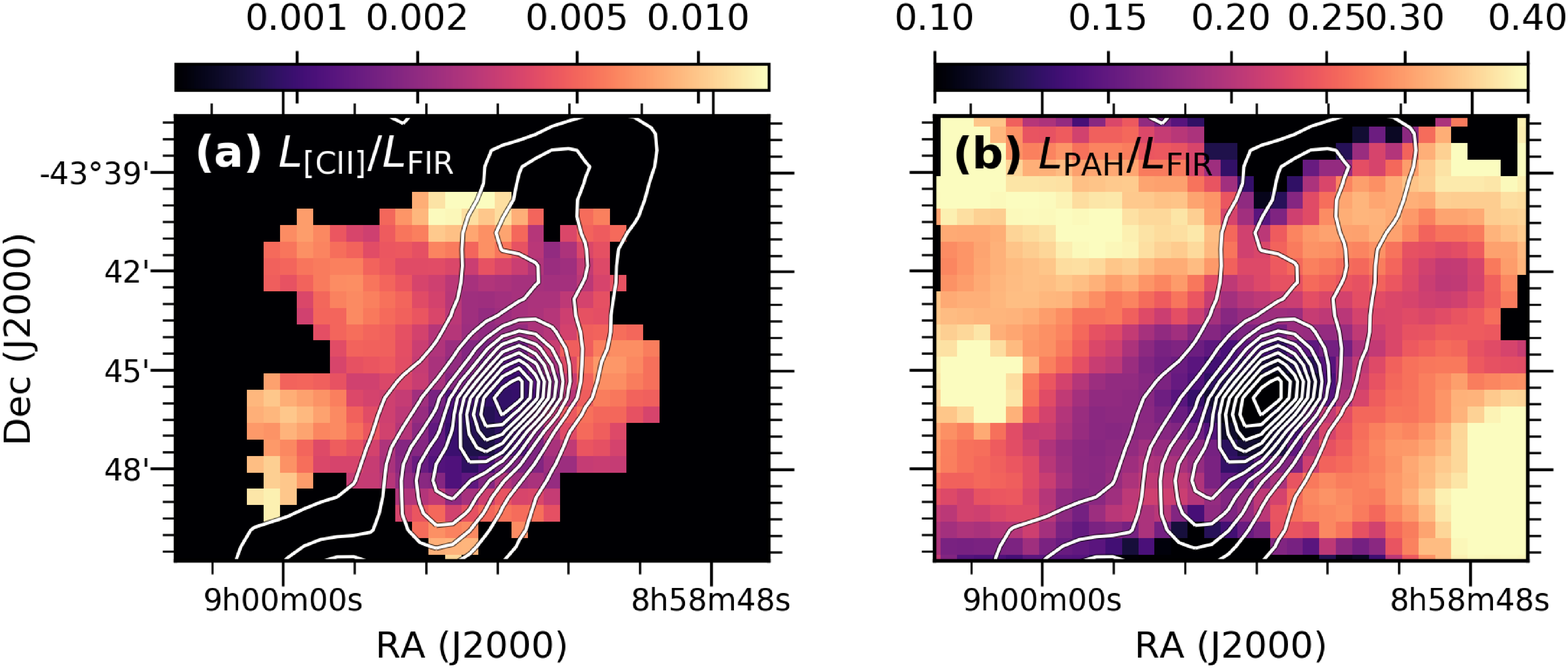}
 \caption{Luminosity ratio maps of (a) $L_\mathrm{[\ion{C}{ii}]}/L_\mathrm{FIR}$
 and (b) $L_\mathrm{PAH}/L_\mathrm{FIR}$ obtained from
 Fig.~\ref{fig:sedfitres}. The $L_\mathrm{vc}$ contours superimposed on the images are
 linearly spaced 10 levels from 0.95 to 9.5 $\mathrm{L_\sun}$}
 \label{fig:PAHabundance_map}
\end{figure*}

\subsection{Enhanced brightness ratio of [CII]/160~$\mu$m}
In Fig.~\ref{fig:correlation}, [\ion{C}{ii}]/160~$\mu$m values for
RCW~36 are systematically higher than those for RCW~38, while
[\ion{C}{ii}]/9~$\mu$m values for RCW~36 are in good agreement with
those for RCW~38. Those results indicate that the PAH emission relative
to the large-grain emission in RCW~36 is higher than that in RCW~38: difference in 
SEDs between RCW~36 and RCW~38.

From the previous two sections, results of
pixel-by-pixel correlation
analyses for RCW~36 suggest that the 160~$\mu$m emission traces the column density of 
dust grains in overall clouds along the line of sight, while [\ion{C}{ii}]
and PAH emissions mainly arise from the far-UV illuminated surface of
clouds. RCW~36 is formed in a filamentary molecular
cloud with H$_2$ column densities ($N_\mathrm{H_2}$) of
$\sim10^{22}$--$10^{23}$~cm$^{-2}$~\citep{Hill2011}. In the outside 
(east-west side) of the filament, $N_\mathrm{H_2}$ is rapidly dropped to
$\sim10^{21}$~cm$^{-2}$: low-column density areas of cold gas and dust
grains. In this situation, the [\ion{C}{ii}] map shows that PDRs are 
formed not only in the filament but also in low-column density areas
because of the bipolar outflow. Therefore, as shown in
Fig.\ref{fig:PAHabundance_map}, higher values of 
$L_\mathrm{[\ion{C}{ii}]}/L_\mathrm{FIR}$ and
$L_\mathrm{PAH}/L_\mathrm{FIR}$ are observed in low-column
density areas ($\tau_\mathrm{d,160} \loa 5\times10^{-2}$) rather than
in the filament.

Based on the picture of the geometry effect, the overall difference in
[\ion{C}{ii}]/160~$\mu$m between RCW~36 and RCW~38 can be
explained by the difference in the total dust-cloud column relative to
the [\ion{C}{ii}] emitting column. Far-UV photons from massive stars in
a filamentary molecular cloud like RCW~36 are likely to leak into
low-dust column density regions perpendicular to the filament, while
those in a clumpy molecular cloud as in the case of RCW~38 are less likely because
stars are surrounded by high-dust column density
regions~\citep{Wolk2006, Kaneda2013, Fukui2016}. In such
low-dust column density regions, the [CII] emitting layer is considered
to be dominated. Therefore, the enhanced brightness ratio of [\ion{C}{ii}]/160
$\mu$m suggests that RCW36 is dominated by far-UV illuminated cloud
surfaces (diffuse PDRs) compared with RCW38.

\section{Conclusions}
Large-area [\ion{C}{ii}]~158~$\mu$m mapping of RCW~36 was conducted to
investigate properties of [\ion{C}{ii}] emission influenced by a
radiative feedback from massive stars formed in the dense filamentary
cloud. The [\ion{C}{ii}] emission overall extends toward the east-west
direction from the peak, which is almost perpendicular to the cold dense
filament direction. There are two prominent arms extending from the
[\ion{C}{ii}] peak to the north-east and south-west directions, which
are spatially in good agreement with shell-like structures of the bipolar
lobe clearly seen in the cold dust ($T_\mathrm{d}\sim30$~K) and PAH
emissions. From a pixel-by-pixel correlation analysis,
the [\ion{C}{ii}] emission spatially correlates with PAH and cold dust
emissions. Therefore, the observed [\ion{C}{ii}] emission dominantly
comes from PDRs. Moreover, we found that the brightness
ratio of [\ion{C}{ii}]/160~$\mu$m for RCW~36 is systematically higher
than that for RCW~38, while that of [\ion{C}{ii}]/9~$\mu$m for RCW~36 is
consistent with RCW~38. The $L_\mathrm{[\ion{C}{ii}]}/L_\mathrm{FIR}$
ratio shows large variation ($10^{-2}$--$10^{-3}$) compared with the
$L_\mathrm{[\ion{C}{ii}]}$/$L_\mathrm{PAH}$ ratio. Given the fact that 
the tight correlation between $L_\mathrm{[\ion{C}{ii}]}/L_\mathrm{FIR}$
and $\tau_\mathrm{d,160}$, the large variation in
$L_\mathrm{[\ion{C}{ii}]}/L_\mathrm{FIR}$ can be explained by that in
the total dust-cloud column relative to the [\ion{C}{ii}] and PAHs
emitting column along each line of sight (geometry effect). Based on the
picture of the geometry effect, the enhanced brightness ratio of
[\ion{C}{ii}]/160~$\mu$m is attributed to the difference in gas
structures where massive stars are formed; unlike RCW~38 where massive
stars are formed inside a clumpy molecular cloud, RCW~36 is formed in a
filamentary molecular cloud and is then dominated by
far-UV illuminated cloud surfaces. Thus, the
difference in large-scale gas structures causes the enhanced brightness
ratio of [\ion{C}{ii}]/160~$\mu$m.

\begin{acknowledgements}
We greatly appreciate all the members of the Infrared Astronomy Group of TIFR
and the staff members of the TIFR Balloon Facility in Hyderabad, India, for their support during balloon flights.
 
We acknowledge support of the Department of Atomic Energy,
Government of India, under project no. 12-R\&D-TFR-5.02-0200
 
The authors gratefully acknowledge the contribution of the anonymous
 referee's comments in improving our manuscript and also thank Dr. Sano
 for providing contours of CO data.
 
 Part of this work is based on observations with AKARI, a JAXA project
 with the participation of ESA. 

 PACS has been developed by a consortium of institutes led by MPE
(Germany) and including UVIE (Austria); KU Leuven, CSL, IMEC (Belgium);
CEA, LAM (France); MPIA (Germany); INAF-IFSI/OAA/OAP/OAT, LENS, SISSA
(Italy); IAC (Spain). This development has been supported by the funding
agencies BMVIT (Austria), ESA-PRODEX (Belgium), CEA/CNES (France), DLR
(Germany), ASI/INAF (Italy), and CICYT/MCYT (Spain). 

 SPIRE has been developed by a consortium of institutes led by Cardiff
University (UK) and including Univ. Lethbridge (Canada); NAOC (China);
CEA, LAM (France); IFSI, Univ. Padua (Italy); IAC (Spain); Stockholm
Observatory (Sweden); Imperial College London, RAL, UCL-MSSL, UKATC,
Univ. Sussex (UK); and Caltech, JPL, NHSC, Univ. Colorado (USA). This
development has been supported by national funding agencies: CSA
(Canada); NAOC (China); CEA, CNES, CNRS (France); ASI (Italy); MCINN
(Spain); SNSB (Sweden); STFC, UKSA (UK); and NASA (USA). 
 This research is supported by JSPS KAKENHI Grant Numbers JP25247020,
 JP18H01252. 
\end{acknowledgements}

\bibliographystyle{aa} 
\bibliography{my_ref.bib} 

\end{document}